\documentclass[english,12pt,aps,prd,a4paper,preprintnumbers,floatfix,nofootinbib,showpacs,superscriptaddress]{revtex4-1}
\usepackage{amsmath}
\usepackage{amstext}
\usepackage{amssymb}
\usepackage{graphicx}
\usepackage{hyperref}
\usepackage{rotating}
\usepackage{gensymb}
\usepackage[normalem]{ulem} 
\usepackage[utf8]{inputenc}
\usepackage{color} 
\usepackage{multirow} 
\usepackage[T1]{fontenc} 
\usepackage{slashed}
\usepackage{caption}
 \usepackage[sort&compress]{natbib} 
\usepackage[english]{babel}
\usepackage{amsmath}
\usepackage{graphicx}
\usepackage{color}
\usepackage{mathrsfs}   
\usepackage{amssymb}
\usepackage{hyperref}
\usepackage{dcolumn}
\usepackage{bm}
\usepackage{graphicx}
\usepackage{mathrsfs}   
\usepackage{slashed}     
\usepackage{url} 
\usepackage{graphicx}
\usepackage{float}
\usepackage{caption}

\definecolor{greenLinks}{rgb}{0, 0.6, 0} 
\definecolor{blueLinks}{rgb}{0, 0, 0.6}
\definecolor{redLinks}{rgb}{0.6, 0, 0}
\definecolor{tempText}{rgb}{0.55, 0.10,0.67}
\definecolor{eprintLinks}{rgb}{0.4, 0.4, 0.4}
\definecolor{journalLinks}{rgb}{0.6, 0, 0}

\newcommand{\MYhref}[3][redLinks]{\href{#2}{\color{#1}{#3}}}%

\usepackage{float}

\def\gsim{\raise0.3ex\hbox{$\;>$\kern-0.75em\raise-1.1ex\hbox{$\sim\;$}}}
\def\lsim{\raise0.3ex\hbox{$\;<$\kern-0.75em\raise-1.1ex\hbox{$\sim\;$}}}

\def\21{$\mathrm{SU(2)_L \otimes U(1)_Y}$ }

\newcommand{\sm}{standard model }

\newcommand{\eVq}  {\text{eV}^2}
 \usepackage[sort&compress]{natbib}
\newcommand{\AddrAHEP}{%
  AHEP Group, Institut de F\'{i}sica Corpuscular --
  C.S.I.C./Universitat de Val\`{e}ncia, Parc Cientific de Paterna.\\
 C/ Catedratico Jos\'e Beltr\'an, 2 E-46980 Paterna (Val\`{e}ncia) - SPAIN}

\begin{document}


\title{ Testing a lepton quarticity flavor theory of neutrino
  oscillations with the DUNE experiment} 

\author{Rahul Srivastava}\email{rahulsri@ific.uv.es}
\author{C. A. Ternes}\email{chternes@ific.uv.es}
\author{M. T{\'o}rtola}\email{mariam@ific.uv.es}  
\author{J.~W.~F.~Valle} \email{valle@ific.uv.es} 
\affiliation{\AddrAHEP}

\begin{abstract}
\vspace{1cm}

  Oscillation studies play a central role in elucidating at least
  some aspects of the flavor problem. 
  Here we examine the status of the predictions of a lepton quarticity
  flavor theory of neutrino oscillations against the existing global
  sample of oscillation data.
  By performing quantitative simulations we also determine the
  potential of the upcoming DUNE experiment in narrowing down the
  currently ill-measured oscillation parameters $\theta_{23}$ and
  $\delta_{\text{CP}}$. We present the expected improved sensitivity
  on these parameters for different assumptions.

 \end{abstract}

\maketitle

\section{Introduction}
\label{sec:intro}

Despite the overwhelming success of the \sm of particle physics, it does
not shed any light on the understanding of the masses and mixings of
quarks and leptons - the so-called flavor problem.
The experimental discovery of neutrino
oscillations~\cite{Kajita:2016cak,McDonald:2016ixn} not only
constitutes the first window into particle physics beyond the Standard
Model, but also exacerbates the challenge posed by the flavor problem.
Indeed, the observed pattern of neutrino oscillation
parameters~\cite{deSalas:2017kay} indicates that leptons are very
different from quarks insofar as the pattern of their charged
current mixing is concerned.

There have been several recent theoretical models proposed in order to
address the flavor problem by incorporating various flavor
symmetries~\cite{babu:2002dz,altarelli:2004za,Morisi:2012fg,Morisi:2013qna,Chen:2015jta,Ma:2016nkf,
  CarcamoHernandez:2017owh,Ma:2017trv,King:2017guk} to account for the
valuable information that comes from oscillation studies.
An alternative approach focusing upon the possible residual CP
symmetries characterizing the neutrino mass matrices, irrespective of
the details of the underlying theory, has also been considered
in~\cite{Chen:2015siy,Chen:2016ica}.

Some of these theoretical
constructions~\cite{Chen:2015jta,Morisi:2013qna} have prompted
dedicated studies confronting their predictions with global neutrino
oscillation
data~\cite{Pasquini:2016kwk,Chatterjee:2017xkb,Chatterjee:2017ilf}.
Here we consider a previously proposed neutrino oscillation
theory. The flavor model construction implements an $A_4$ flavor
symmetry as well as lepton quarticity
symmetry~\cite{CentellesChulia:2017koy}. The latter correlates dark
matter stability with the predicted Dirac nature of
neutrinos~\cite{Chulia:2016ngi}.

 While this is an interesting connection in itself, leading to a
  viable dark matter scenario, it leads to novel neutrino predictions,
  for example the presence of neutrinoless quadruple beta decay
  ($0 \nu 4 \beta$) signal in the absence of neutrinoless double beta
  decay ($0 \nu 2 \beta$)~\cite{Hirsch:2017col}.
  Considering that Majorana neutrinos have so far remained
  elusive~\cite{Agostini:2017iyd,Albert:2014awa,Alduino:2017ehq} the
  possibility that the quadruple beta decay might exist on its
  own~\cite{Heeck:2013rpa} is especially intriguing and has already
  been subject to a dedicated experimental search by the NEMO
  collaboration~\cite{Arnold:2017bnh}.

Apart from these interesting features of the model, which arise from
the quarticity symmetry, the model has other novel features owing to
the presence of $A_4$ flavor symmetry.  Thanks to the latter, the tree
level dimension-4 Dirac mass terms for the neutrinos are
forbidden. However, the $A_4$ symmetry allows us to generate
\textit{seesaw-induced} small neutrino masses.
In addition, the model predicts a successful generalized ``golden''
Bottom-Tau unification
formula~\cite{Morisi:2011pt,King:2013hj,Morisi:2013eca,Bonilla:2014xla},
as well as definite predictions for neutrino oscillations.
For example, the scheme leads to normal neutrino mass ordering.  It
also leads to a strong correlation between the two currently
ill-measured oscillation parameters, namely the leptonic CP phase
$\delta_{\text{CP}}$ and the mixing angle $\theta_{23}$. This correlation in
turn implies that CP must be significantly violated in neutrino
oscillations, with the atmospheric angle $\theta_{23}$ lying in the
second octant.

Owing to the precise predictions made by the model, it constitutes an
ideal candidate to be probed at the forthcoming long baseline
oscillation experiments aimed at measuring $\delta_{\text{CP}}$ and
$\theta_{23}$, such as DUNE.
Here we scrutinize the neutrino oscillation predictions 
  obtained in the model against the latest available global
neutrino oscillation study~\cite{deSalas:2017kay}  as well as
  the future discriminating power of the DUNE experiment.
  Our strategy here is then, given the current measurements of the
  four ``well determined'' oscillation parameters, i.e. the mass
  splittings characterizing solar and atmospheric oscillations plus
  the two mixing angles $\theta_{12}$ and $\theta_{13}$, to determine
  the potential of the upcoming DUNE experiment in narrowing down the
  still poorly measured parameters
  $\theta_{23}$-$\delta_{\text{CP}}$.
We do this for the general ``unconstrained'' three-neutrino
oscillation paradigm, as well for our ``constrained'' scenario in
which the model predictions are taken into account.
From our results we conclude that substantial improvements are to be
expected.

\section{Current status of the model}
\label{sec:glob}

In order to define our goal we first determine the current status of
the neutrino oscillation parameters within the model, by taking into
account the latest global analysis.
We provide an improved update of the model neutrino oscillation
predictions~\cite{CentellesChulia:2017koy}, originally tested against
our previous oscillation global fit presented in~\cite{Forero:2014bxa}
assuming only the one-dimensional 3$\sigma$ intervals.
Here we confront the model with the new results~\cite{deSalas:2017kay}
and use the more complete $\chi^2$-distributions.
In order to do so, we generated many points consistent with the model
predictions.  The latter are obtained by first randomly varying all
the free parameters such as Yukawa couplings and new scalar field vevs
over their allowed theoretical ranges. The points thus obtained are
then tested for their compatibility with the currently well measured
observables, such as quark and charged lepton masses~\footnote{The
  constraints coming from neutrino oscillation parameters are not
  imposed at this point. However, we do impose the cosmological limit
  on sum of neutrino masses \cite{Lattanzi:2016rre}. }. For more
details on the model predictions see \cite{CentellesChulia:2017koy}.
Only points within the 3$\sigma$ range of these parameters are
retained as genuine points.
Next, using the results from the global fit to neutrino oscillations
in Ref.~\cite{deSalas:2017kay}, we assign to each of those points a
$\chi^2$-value that quantifies their agreement with most recent data.
 Given the negligible effect of solar parameters in DUNE we have
  simply selected those oscillation parameter sets with solar
  parameters within their allowed 3$\sigma$ region, as derived in
  \cite{deSalas:2017kay}. 
  The resulting 4-dimensional $\chi^2$-maps are minimized over
    $\Delta m_{31}^2$ and $\sin^2 \theta_{13}$, leading to the
      final distribution as a function of the parameters of
      interest, $\theta_{23}$ and $\delta_{\text{CP}}$.
\begin{figure}[t!]
 \centering
        \includegraphics[width=0.5\textwidth]{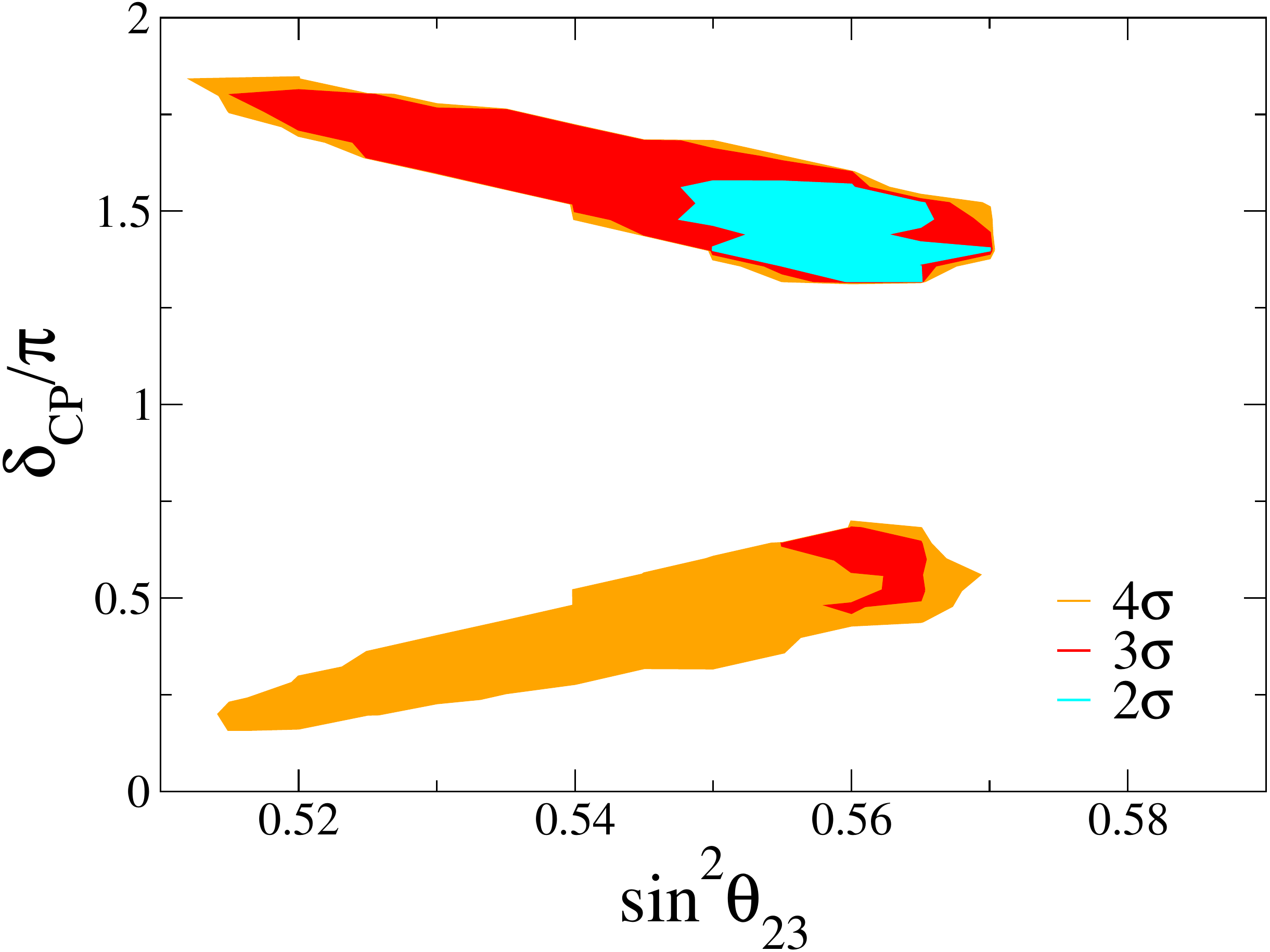}
         \captionsetup{justification=raggedright}
        \caption{Allowed regions at 2, 3 and 4$\sigma$ in the
          plane $\theta_{23}$-$\delta_{\text{CP}}$ within the model,
          given the current global neutrino oscillation analysis.}
	\label{fig:MODEL-status}
\end{figure}

The results for the current status of the model are presented in
Fig.~\ref{fig:MODEL-status}.  One finds that, thanks to the new
oscillation data, the predicted regions have shrunk significantly with
respect to those in Ref.~\cite{CentellesChulia:2017koy}, so that CP
phase values $\delta_{\text{CP}} \leq \pi $ are allowed only at
3$\sigma$.
Indeed, one sees that, at 3$\sigma$ confidence level, one of the two
allowed branches has nearly disappeared.
We find that the new 4$\sigma$ regions are roughly similar to the old
3$\sigma$ regions, and there are no points surviving at the 1$\sigma$
confidence level.
 Note that neither the current best fit point nor the local
  minimum from the global neutrino oscillation fit
  in~\cite{deSalas:2017kay} lies within the parameter region predicted
  by the model. As a result, if new data confirm the current best fit
  point, the local minimum or nearby values, the model would be
  strongly disfavored.

\section{Simulation of the DUNE experiment}
\label{sec:simulation}

We simulate the DUNE experiment using the GLoBES
package~\cite{Huber:2004ka,Huber:2007ji} and the auxiliary
file~\cite{Alion:2016uaj} used to produce the plots in
Ref.~\cite{Acciarri:2015uup}.
In our simulation DUNE is running 3.5 years in the neutrino mode and
another 3.5 years in the anti-neutrino mode. Using its 80 GeV beam with
1.07 MW beam power and the 40~kt far detector, this gives an exposure of
300 kton-MW-years, which corresponds to $1.47\times 10^{21}$ protons
on target (POTs) each year.

We consider the disappearance channels for neutrinos and
anti-neutrinos, as well as the appearance channels.  We also simulate
the backgrounds, taking into account several sources of errors in our
simulation, where we assign a 2\% error on the signals in the
appearance channels and 5\% in the disappearance channels, as
indicated in the studies performed by the DUNE
Collaboration~\cite{Acciarri:2015uup}.
Likewise, we have implemented backgrounds ranging between
  5\% and 20\%. These include misinterpretation of neutrinos as
  antineutrinos and vice-versa, contamination of electron neutrinos
  and antineutrinos in the beam, misinterpretation of muon as electron
  neutrinos, as well as the appearance and misinterpretation of tau
  neutrinos and neutral current interactions.

  Here we are mainly interested in the currently poorly determined
  oscillation parameters $\sin^2\theta_{23}$ and $\delta_{\text{CP}}$. 
  Therefore, in order to simulate the future event rate in DUNE we fix
  the rest of the parameters to their best fit values reported in
  \cite{deSalas:2017kay}.
  Then, in the statistical analysis performed to determine the
    DUNE sensitivity, we marginalize over $\theta_{13}$,
    $\theta_{12}$, $\Delta m_{31}^2$ and $\Delta m_{21}^2$ within
    their 1$\sigma$-ranges, see Table~\ref{tab:params-margs}.
    Concerning the parameters of interest, we generate future DUNE
    data by assuming several pairs of
    $(\theta_{23}^\text{true},\delta_{\text{CP}}^\text{true})$.
    For each set of reconstructed parameters
    $(\theta_{23},\delta_{\text{CP}})$ we calculate the
    $\chi^2$-function, given as
\begin{equation}
 \chi^2(\theta_{23},\delta_{\text{CP}})=
 \min_{\theta_{1j}, \Delta m_{j1}^2,\vec{\alpha}}\sum_\text{channels} 2\sum_n\left[ N_n^{\text{test}}- N_n^{\text{dat}} +
 N_n^{\text{dat}}\log\left(\frac{N_n^{\text{dat}}}{N_n^{\text{test}}}\right)\right] 
 + \sum_i \left(\frac{\alpha_i}{\sigma_i}\right) ,
 \label{main-chi2}
\end{equation}
where $\theta_{1j}, \Delta m_{j1}^2$ (j=2,3) denote the four
well-measured oscillation parameters.
Here $N_n^\text{dat}$ corresponds to the simulated event number in the
$n$-th bin obtained with $\theta_{23}^\text{true}$ and
$\delta_{\text{CP}}^\text{true}$.
$N_n^{\text{test}}$ is the event
number in the $n$-th bin associated to the parameters
$(\theta_{23},\delta_{\text{CP}})$ and $\alpha_i$ and $\sigma_i$ are
the nuisance parameters and their corresponding standard deviations,
respectively.
  Although not explicitly shown, note that
  $N_n^{\text{test}}$ also depends on
  $\vec\alpha$.
\begin{table}[t!]\centering
  \catcode`?=\active \def?{\hphantom{0}}
   \begin{tabular}{|l|c|c|}
    \hline
    parameter & best fit value & relative error
    \\
    \hline
    $\Delta m^2_{21}$& $7.56\times 10^{-5}\eVq$&2.5\%\\  
    $\Delta m^2_{31}$&  $2.55\times 10^{-3}\eVq$&1.6\%\\
    $\sin^2\theta_{13}$ & 0.02155&3.9\%\\
    $\sin^2\theta_{12}$ & 0.321&5.5\%\\
    \hline
     \end{tabular}
          \captionsetup{justification=raggedright}
          \caption{ Best fit values and 1$\sigma$ relative
            uncertainties for the better determined neutrino
            oscillation parameters from \cite{deSalas:2017kay}.}
     \label{tab:params-margs} 
\end{table}

\section{Results}
\label{sec:results}

In this section we present the main results of the analyses which we
have performed in order to test the neutrino oscillation model in
question.
We start in Sec.~\ref{sec:DUNE-run} by performing an unconstrained
DUNE sensitivity analysis for seven years of run time, assuming 3.5
years runs in both neutrino and anti-neutrino mode. In this analysis
we have assumed that $\theta^{\text{true}}_{23}$ and
$\delta^{\text{true}}_{\text{CP}}$ lie within the 1$\sigma$ region
obtained in the recent neutrino oscillation global
fit~\cite{deSalas:2017kay}.  This analysis is performed in order to
quantify the projected sensitivity of the DUNE experiment given the
current status of these parameters, and is completely
model-independent.

In Sec.~\ref{sec:cons}, we present the expected sensitivity on the
currently ill-measured parameters after seven years running of DUNE,
assuming that $\theta^{\text{true}}_{23}$ and
$\delta^{\text{true}}_{\text{CP}}$ lie in the range predicted by the
model. In this analysis we have taken into account only the model
prediction for these parameters and have not taken into account the
current global oscillation fit.
Finally, in Sec.~\ref{sec:cons-global} we perform a combined analysis
of the expected DUNE sensitivity taking into account, as input, both
the range predicted by the model as well as the current oscillation
global fit. The different analyses are performed in order to highlight
the discriminating power of DUNE in various scenarios of interest,
both from the model point of view as well from that of the current
global fit.

\subsection{Model-independent DUNE sensitivity}
\label{sec:DUNE-run}

\begin{figure}[!h]
 \centering
      \includegraphics[width=0.8\textwidth]{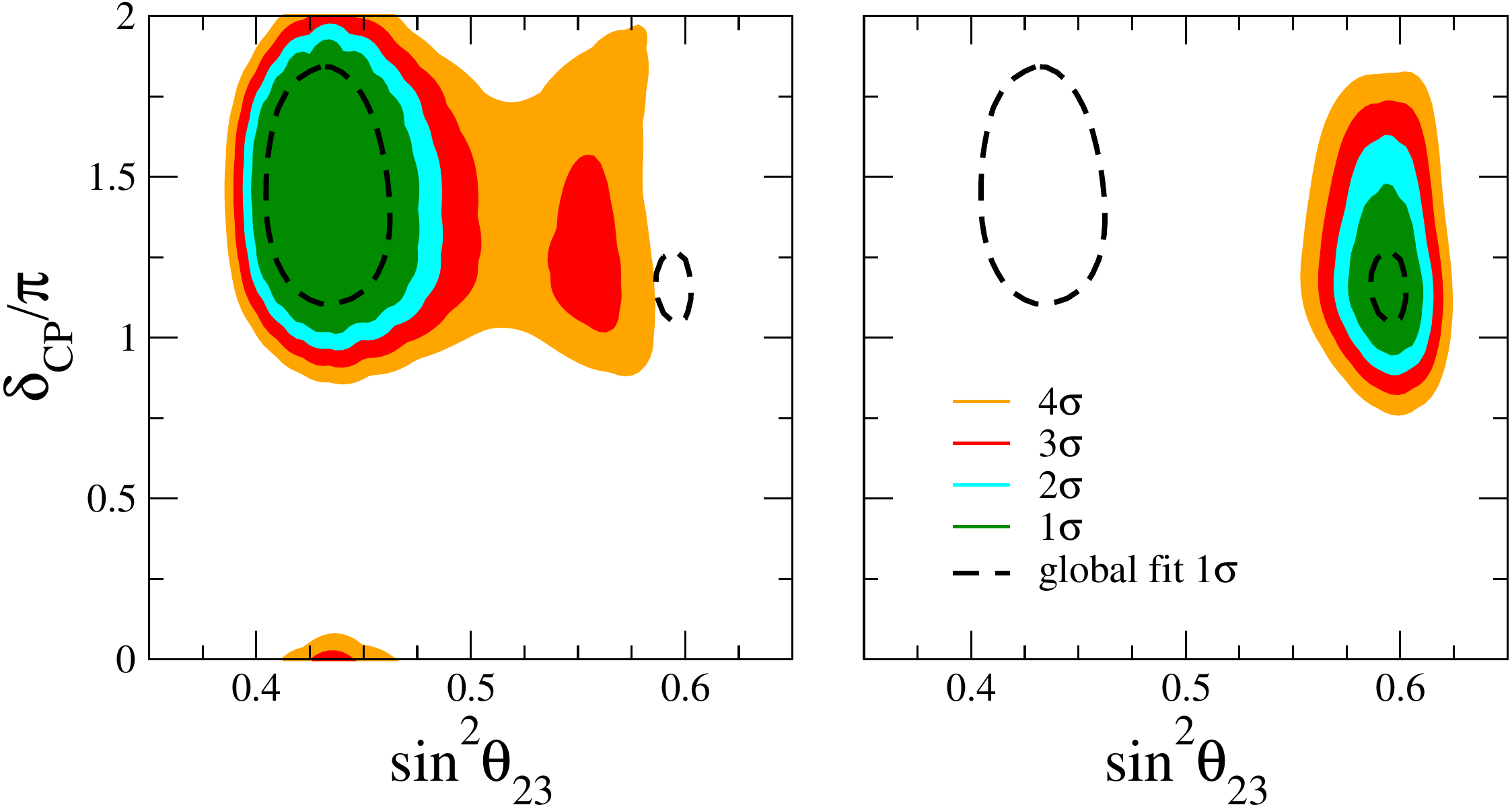}
       \captionsetup{justification=raggedright}
       \caption{ Potential of DUNE running 7 years in narrowing down
         the currently ill-measured parameters
         $\theta_{23}$-$\delta_{\text{CP}}$. The left (right) panel
         only assumes that $\theta_{23}^\text{true}$ lies in the lower
         (upper) octant. Both $\theta_{23}^\text{true}$ and
         $\delta_{\text{CP}}^\text{true}$ are allowed to vary in their
         allowed 1$\sigma$-regions. The dashed black line is the
         1$\sigma$ region from the global fit performed
         in~\cite{deSalas:2017kay}.  See text for more details.}.
	\label{fig:DUNE+GLOBALFIT}
\end{figure}

As explained above, in this section we study the sensitivity of DUNE
to $\theta_{23}$ and $\delta_{\text{CP}}$, taking into account the
current status of neutrino oscillations as reported
in~\cite{deSalas:2017kay}.
Assuming the true oscillation parameters to be the current best fit
values would be too strong an assumption.
We have therefore decided to vary $\theta_{23}^\text{true}$ and
$\delta_\text{CP}^\text{true}$ within their 1$\sigma$ ranges for two
degrees of freedom (d.o.f.), indicated by the dashed black lines in
Fig.~\ref{fig:DUNE+GLOBALFIT}.
We have performed this analysis separately for the values in the lower
and the upper octant of the atmospheric angle. For this we have
defined
\begin{equation}
 \chi^2_{1\sigma}(\theta_{23},\delta_{\text{CP}}) = \min_{\theta_{23}^\text{true},\delta_{\text{CP}}^\text{true}} \chi^2(\theta_{23},\delta_{\text{CP}}),
\end{equation}
where $(\theta_{23}^\text{true},\delta_{\text{CP}}^\text{true})$ run
first over all the values allowed in the lower octant, and later over
all those allowed in the upper octant. Here
$\chi^2(\theta_{23},\delta_{\text{CP}})$ is the function given in
Eq.~\ref{main-chi2}.  

The results of this minimization can be seen in
Fig.~\ref{fig:DUNE+GLOBALFIT}, where we plot the 1$\sigma$, 2$\sigma$,
3$\sigma$ and 4$\sigma$ allowed regions for 2 d.o.f in the
$\sin^2\theta_{23}$--$\delta_{\text{CP}}$ plane.
The left (right) panel corresponds to the analysis assuming
$\theta_{23}^\text{true}$ to lie in the lower (upper) octant. 
At the moment, the lower octant is preferred by the global oscillation
data, and therefore there are much more points in this region,
resulting in a bigger region in our plot.

Notice that in the left panel, the degenerate solution in the second
octant appears only at the 3$\sigma$ confidence level.
Conversely, if the true value of the atmospheric mixing angle lies in
the small region in the upper octant (see right panel of
Fig.~\ref{fig:DUNE+GLOBALFIT}), the degenerate first-octant solution
would be ruled out at more than 4$\sigma$. Maximal mixing is
disfavored at more than 3$\sigma$ (5$\sigma$) for
$\theta_{23}^\text{true}$ in the lower (upper) octant.
In both cases, values of $\delta_\text{CP}\approx 0.5\pi$ would be
excluded with very high significance.

\subsection{Testing the model with DUNE}
\label{sec:cons}

\begin{figure}[b]
 \centering
        \includegraphics[width=0.5\textwidth]{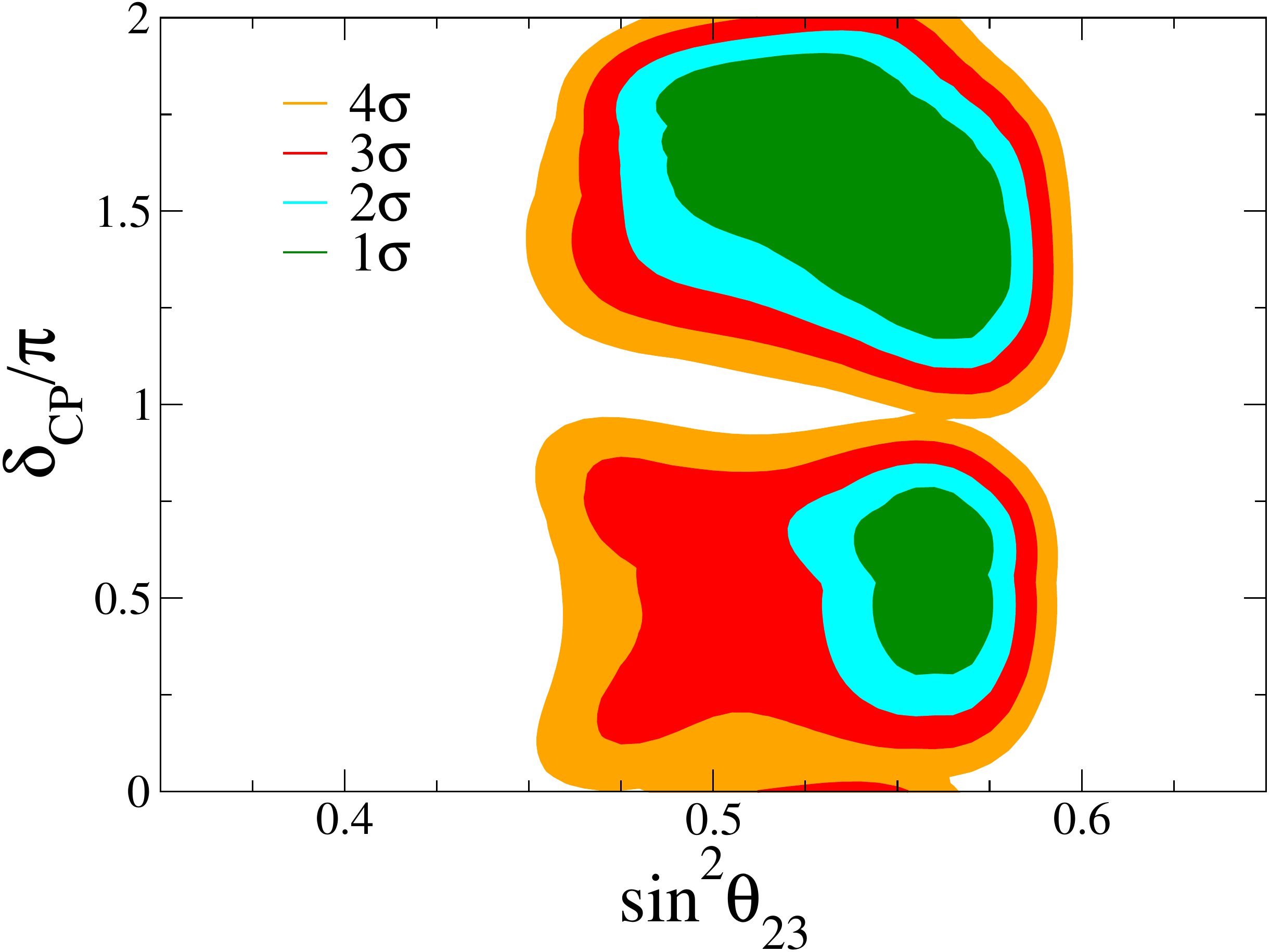}
         \captionsetup{justification=raggedright}
         \caption{ DUNE sensitivity to the ($\sin^2\theta_{23}$,
           $\delta_\text{CP}$) parameter region predicted by the model
           consistent with the current global fit at 3$\sigma$.}
	\label{fig:DUNE+MODEL}
\end{figure}

In order to quantify the sensitivity of DUNE to test the model
predictions, we now perform a simulation of DUNE suited to the model of
interest.
Our procedure will not depend on any input from global neutrino
oscillation fits.  This means we assume the model prediction for
$\theta_{23}$ and $\delta_{\text{CP}}$ to be the true values used to
generate DUNE data.  As in the last section, we define the $\chi^2$
function as
\begin{equation}
 \chi^2_{\text{DUNE+model}}(\theta_{23},\delta_{\text{CP}}) = \min_{\theta_{23}^\text{true},\delta_{\text{CP}}^\text{true}} \chi^2(\theta_{23},\delta_{\text{CP}})\,.
\end{equation}
In this case,
$(\theta_{23}^\text{true},\delta_{\text{CP}}^\text{true})$ are not the
values from the 1$\sigma$ regions of~\cite{deSalas:2017kay}, but
include, instead, all  the points predicted by the model and
consistent at 3$\sigma$ with the current global fit, see
Fig.~\ref{fig:MODEL-status}. 
The resulting regions corresponding to 1$\sigma$ to 4$\sigma$
confidence level for 2 d.o.f.  are presented in
Fig.~\ref{fig:DUNE+MODEL}.
These results correspond to the case where only the model
predictions are taken into account. 
In this case one finds that, by themselves, model predictions plus
DUNE data would not suffice to determine the octant of the atmospheric
angle or a unique preferred range for the CP phase, at least not
  for all parameter choices~\footnote{Notice that, ideally, only one
    set of $(\theta_{23},\delta_\text{CP})$ would be realized in
    nature as the true value while, to be conservative, here we
    marginalize over all neutrino oscillation parameters possible in
    the model.}.

By including valuable information on the current status of
   global fits to neutrino  oscillations  one can sharpen   the expected DUNE sensitivity to the oscillation parameters, beyond
  the results of Fig.~\ref{fig:DUNE+MODEL}. This is done in the next
  section.

\subsection{Testing the model with DUNE: the global picture}
\label{sec:cons-global}

\begin{figure}[!h]
 \centering
        \includegraphics[width=0.56\textwidth]{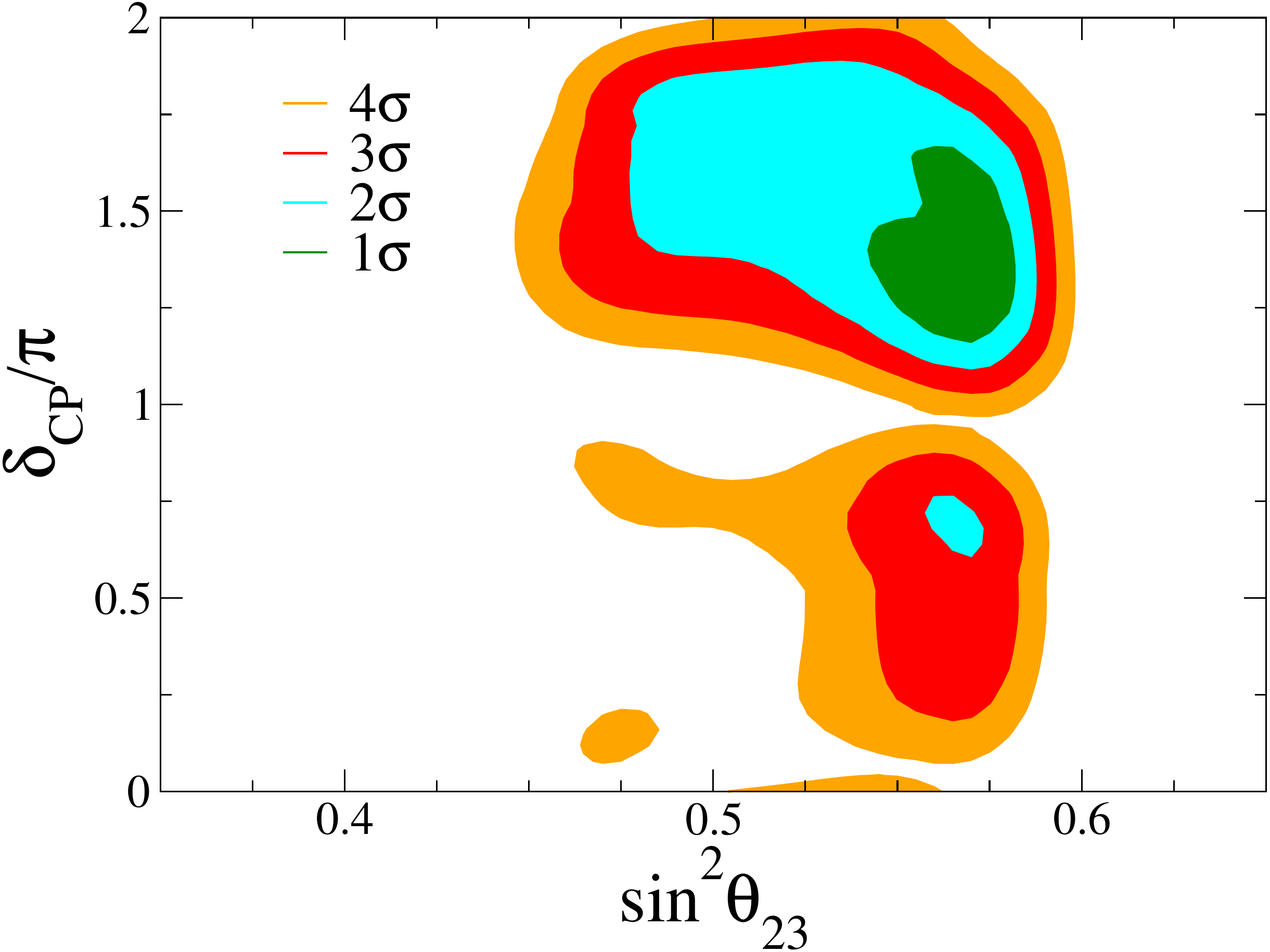}
         \captionsetup{justification=raggedright}
         \caption{ DUNE sensitivity to the ($\sin^2\theta_{23}$,
           $\delta_\text{CP}$) parameter region predicted by the
           model, taking into account the constraints from the current
           global neutrino oscillation fit in \cite{deSalas:2017kay}.}
	\label{fig:DUNE+MODEL+GLOBALFIT}
\end{figure}

In order to better quantify the sensitivity of DUNE to test the model
predictions we now perform a ``constrained global neutrino oscillation
fit'' suited to the model of interest. We do this by combining the
DUNE simulation with the global fit to neutrino oscillations in
Ref.~\cite{deSalas:2017kay} in the context of the lepton quarticity
flavor model under study.
In order to combine the results of the DUNE simulations performed here
with the global analysis of neutrino oscillation data we simply sum
the $\chi^2$ function defined in Sec.~\ref{sec:cons} with the $\chi^2$
grid obtained in the global fit to neutrino oscillations in
Ref.~\cite{deSalas:2017kay},
\begin{equation}
 \chi^2_{\text{tot}}(\theta_{23},\delta_{\text{CP}}) = \chi^2_{\text{DUNE+model}}(\theta_{23},\delta_{\text{CP}}) + \chi^2_{\text{fit}}(\theta_{23},\delta_{\text{CP}}).
\end{equation}
The results are presented in Fig.~\ref{fig:DUNE+MODEL+GLOBALFIT},
where we plot the regions allowed at 1$\sigma$ to 4$\sigma$ confidence
level for 2 d.o.f.
One sees that by combing all the relevant information the regions
shrink with respect to those of Fig.~\ref{fig:DUNE+MODEL}, since the
global fit to current neutrino data disfavors $\delta_{\text{CP}}$ in
the range $[0,\pi]$. This result is shown in the corresponding panel
of Fig.~8 in \cite{deSalas:2017kay}.
One sees that, by properly taking into account the current knowledge
of neutrino oscillation parameters and the model under consideration,
one concludes that DUNE will determine rather well the CP phase at the
1$\sigma$ level, excluding CP-conserving scenarios at more than
3$\sigma$.
One sees also that, at the 1$\sigma$ level, the second octant would be
singled out. Besides that, the status of maximal mixing would worsen
in comparison to Fig.~\ref{fig:DUNE+MODEL}, as a consequence of the
recent global fit results, although it would still remain allowed at
the 2$\sigma$ level for certain consistent model parameter choices.
As commented before, here we are marginalizing over a large set of
true oscillation parameters and therefore, the real sensitivity given
in Fig.~\ref{fig:DUNE+MODEL+GLOBALFIT} would still be too
conservative.

\section{Summary and discussion}
\label{sec:summary-discussion}

Neutrino oscillation studies may play a key role in elucidating major
aspects of the flavor problem.
Here we have provided a quantitative study of the status of the
predictions of a lepton quarticity flavor theory of neutrino
oscillations.
Thanks to the assumed flavor symmetry, the model explains the small
neutrino masses as a result of a variant of the seesaw mechanism
leading to Dirac neutrinos.
Due to quarticity, the model has a viable dark matter candidate
stabilized by the Diracness of neutrinos.
In addition, it leads to a successful ``golden'' Bottom-Tau
unification formula, as well as definite predictions for neutrino
oscillations, first studied in~\cite{CentellesChulia:2017koy}.
  Here we have reexamined the consistency of neutrino oscillation
  model predictions in view of the latest global sample of neutrino
  oscillation data~\cite{deSalas:2017kay}.
  One finds that the model predicts normal neutrino mass ordering, and
  significant violation of CP in neutrino oscillations, with the
  atmospheric angle $\theta_{23}$ lying in the second octant. Our
  results are given in Fig.~\ref{fig:MODEL-status}.
By performing dedicated simulations we have also determined
    the potential of future DUNE data in further restricting the
    currently ill-measured oscillation parameters, $\theta_{23}$ and
    $\delta_{\text{CP}}$.
    Fig.~\ref{fig:DUNE+GLOBALFIT} illustrates the resulting
    sensitivity for the ``unconstrained'' model-independent case,
    assuming the true $\theta_{23}$ and $\delta_\text{CP}$ parameters
    to lie within the 1$\sigma$ region obtained from the recent global
    fit to neutrino oscillations as given in~\cite{deSalas:2017kay}.
    By taking into account not only the information from the current
    neutrino oscillation global fit results but also the specific
    model predictions, we have shown that DUNE data should
    unambiguously single out the second octant of $\theta_{23}$ and
    exclude values of $\delta_{CP}$ below $\pi$ at the 1$\sigma$
    level, as seen in Fig.~\ref{fig:DUNE+MODEL+GLOBALFIT}.
    Finally we stress that, as already mentioned, by marginalizing
    over a large set of potential true oscillation parameter values,
    we are being conservative in our estimate of the improved
    sensitivity of DUNE to test the model under consideration.

\begin{acknowledgments}

  Work supported by the Spanish grants FPA2014-58183-P,  FPA2017-85216-P and
  SEV-2014-0398 (MINECO), PROMETEOII/2014/084 and GV2016-142 grants
  from Generalitat Valenciana.
MT is also supported by a Ram\'{o}n y Cajal contract (MINECO).

\end{acknowledgments}


\bibliographystyle{bib_style_T1}

\end{document}